\documentclass[12pt]{article}

\usepackage{rotating}

\usepackage{multirow}
\usepackage{amsmath,amssymb,graphicx, graphics, color,verbatim} 
\usepackage{cite}
\usepackage{latexsym}  
\usepackage{cancel}
\usepackage{hyperref}

\usepackage{enumerate} 

\usepackage[normalem]{ulem} 

\pdfoutput=1
\usepackage{epsf}
\usepackage{pstricks} 
\usepackage{comment}

\usepackage{array}

\newcommand{\centeron}[2]{{\setbox0=\hbox{#1}\setbox1=\hbox{#2}\ifdim
\wd1>\wd0\kern.5\wd1\kern-.5\wd0\fi \copy0
\kern-.5\wd0\kern-.5\wd1\copy1\ifdim\wd0>\wd1
                                   \kern.5\wd0\kern-.5\wd1\fi}}
\newcommand{\ltap}{\>\centeron{\raise.35ex\hbox{$<$}}
                           {\lower.65ex\hbox{$\sim$}}\>}
\newcommand{\gtap}{\>\centeron{\raise.35ex\hbox{$>$}}
                           {\lower.65ex\hbox{$\sim$}}\>}

\newcommand\ZZ{\hbox{\zfont Z\kern-.4emZ}}
\font\zfont = cmss10 

\newcommand{\fref}[1]{Fig.\ \ref{f.#1}}
\newcommand{\eref}[1]{Eq.\ (\ref{e.#1})}

\newcommand{\aref}[1]{Appendix \ref{a.#1}}
\newcommand{\sref}[1]{Section \ref{s.#1}}
\newcommand{\ssref}[1]{Section \ref{ss.#1}}
\newcommand{\sssref}[1]{Section \ref{sss.#1}}
\newcommand{\cref}[1]{Chapter \ref{c.#1}}
\newcommand{\tref}[1]{Table \ref{t.#1}}

\newcommand{\ba}{\begin{array}}
\newcommand{\ea}{\end{array}}

\newcommand{\beq}{\begin{eqnarray}}
\newcommand{\eeq}{\end{eqnarray}}
\newcommand{\beqs}{\begin{eqnarray*}}
\newcommand{\eeqs}{\end{eqnarray*}}

\newcommand{\bal}{\begin{align}} 
\newcommand{\eal}{\end{align}}

\def\bi{\begin{itemize}}
\def\ei{\end{itemize}}
\def\ben{\begin{enumerate}}
\def\een{\end{enumerate}}
\def\bc{\begin{center}}
\def\ec{\end{center}}
\def\bt{\begin{table}}
\def\et{\end{table}}
\def\btb{\begin{tabular}}
\def\etb{\end{tabular}}

\def\gev{\, {\rm GeV}}

\def\mass2{mass${}^2$}

  \newcommand{\ww}{
$W^+W^-$
}

\newcommand{\ifb}{\mathrm{fb}^{-1}}

\begin{document}
\bibliographystyle{unsrt}
\begin{titlepage}
\begin{flushright}
\small{YITP-SB-14-14}
\end{flushright}

\vskip2.5cm
\begin{center}
\vspace*{5mm}
{\huge \bf Natural SUSY in Plain Sight}
\end{center}
\vskip0.2cm

\begin{center}
{\bf David Curtin, Patrick Meade, Pin-Ju Tien}

\end{center}
\vskip 8pt

\begin{center}
{\it C. N. Yang Institute for Theoretical Physics\\ Stony Brook University, Stony Brook, NY 11794.
\\}
\vspace*{0.3cm}

\vspace*{0.1cm}

{\tt 
david.curtin@stonybrook.edu,
patrick.meade@stonybrook.edu,
pin-ju.tien@stonybrook.edu}
\end{center}

\vglue 0.3truecm

\begin{abstract}

The basic principle of naturalness has driven the majority of the LHC program, but so far all searches for new physics beyond the SM have come up empty. On the other hand, existing measurements of SM processes contain interesting anomalies, which allow for the possibility of new physics with mass scales very close to the Electroweak Scale. 
In this paper we show that SUSY could have stops with masses ~$\mathcal{O}(200)$ GeV based on an anomaly in the \ww cross section, measured by both ATLAS and CMS at 7 and 8 TeV.   In particular we show that there are several different classes of stop driven scenarios that not only evade all direct searches, but improve the agreement with the data in the SM measurement of the \ww cross section.

\end{abstract}

\end{titlepage}

\section{Introduction}
\label{s.intro} \setcounter{equation}{0} \setcounter{footnote}{0}

The impressive performance of the LHC has thrust theoretical physics into a state of some confusion. The discovery by ATLAS and CMS of the Higgs boson~\cite{Higgs125Combined}, or something very much like it, is an unparalleled triumph.  That being said, it also brings the naturalness and hierarchy problems to the fore. We now have to directly confront the possibility that a fundamental scalar has been discovered in nature.  In general, any weakly coupled solution of the hierarchy problem should feature new states below the TeV scale. Unfortunately, no such new states have been discovered so far by either ATLAS or CMS~\cite{moriondEW, moriondQCD}.

Supersymmetry (SUSY) is the most theoretically well-motivated and calculable solution to the hierarchy problem. However, it is this very calculability which naively places it under stronger tension than most other potential solutions. This is because the minimal implementation of SUSY, the MSSM,  predicts the Higgs quartic coupling solely within the IR sector of the theory. While this predictive nature of the MSSM is one of its more desirable features, accounting for the exact mass of the Higgs discovered by ATLAS and CMS requires radiative corrections to the quartic coupling from particles within the MSSM.  The dominant radiative contribution comes from the stops, and a 125 GeV Higgs mass naively requires stops above a TeV. This can easily be accommodated within the MSSM but somewhat counteracts the supersymmetric solution to the hierarchy problem, since the same particles which give radiative contributions to the quartic term in the Higgs potential also cancel its quadratic divergences. This tension, with heavy stops required for a heavy Higgs but light stops required for naturalness, is the so-called ``little hierarchy problem'' of the MSSM. 

There are many model building solutions to the little hierarchy problem within SUSY. Two important examples of theories which generate new Higgs quartic contributions without heavy stops are the NMSSM/$\lambda$SUSY~\cite{nmssm,lambdasusy} and additional D-term contributions~\cite{dterms}. In these models, SUSY can in principle be fully natural, solving the hierarchy problem without fine-tuning, provided that the stops are sufficiently light. This has motivated an extensive LHC program at both ATLAS and CMS in an attempt to cover all possibilities to search for light stops~\cite{stoptotopsearches, otherstopsearchesATLAS, otherstopsearchesCMS,stoptocharmsearches,Aad:2014qaa,ATLAS7stop1}. This logic also extends to other BSM models that solve the hierarchy problem, with both major LHC collaborations~\cite{generictoppartnersearches} working to pin down generic top partners~\cite{meadereece,tophunter}. Despite these efforts, no $3^\mathrm{rd}$ generation partners of SM particles have been found, and lower limits on the masses of particles potentially responsible for naturalness are becoming uncomfortably stringent~\cite{stoptotopsearches, otherstopsearchesATLAS, otherstopsearchesCMS,stoptocharmsearches,Aad:2014qaa,ATLAS7stop1,generictoppartnersearches}.

Given these negative results it is especially important to understand where new physics may have been missed. Of course, it is always possible that new particles are ``just around the corner'' at higher mass scales, but naturalness prompts us to look for lower-lying hiding places. A remarkable possibility is that new physics could still be very close to the electroweak scale.  Searches are typically based on being able to maximally separate new physics from SM backgrounds. However, if new physics is very close to the EW scale it becomes difficult to disentangle and searches lose their sensitivity. Related to this is the even more interesting possibility that new physics already contaminates measurements of SM processes. Hiding stops at low masses has been investigated by many groups in the past~\cite{hidinglightstops}.  Particular attention has been paid to the idea that stops could be at the same mass as the top quarks, or that they could decay via R-parity violation into a jet-rich final state. In both of these scenarios the stop is very difficult to find. The absence of any anomalies means bounds are set by living within the error bars of current measurements.

In~\cite{Curtin:2012nn} it was pointed out that not only could new physics be hiding in searches, but based on existing LHC measurements it could in certain cases \emph{improve} the fit to the data, compared to the SM alone. The work of~\cite{Curtin:2012nn} was based on the  \ww cross section as measured by both ATLAS~\cite{ATLAS:2012mec} and CMS~\cite{CMS:2012tbb} at 7 TeV and with low luminosity at 8 TeV by CMS~\cite{CMS:2012daa}. Both experiments observed a total cross section $\sim 15-20\%$ above the SM expectation, disagreeing with the SM at the $1-2\sigma$ level individually, with a combined significance of about $3\sigma$. Furthermore, the excess seems to be concentrated near the center of the kinematic distributions at moderate $p_T$ and invariant masses, while the tails are very well modeled by the SM. These shape differences, apart from raising the significance of the excess, could be suggestive of additional kinematically distinct contributions to the $\ell \ell + \mathrm{MET}$ final state in which the \ww cross section is measured. 

In addition to the anomalies in the SM measurements, the control region for $h\rightarrow$\ww with 0-jets is also higher than expected for run I~\cite{atlashwwdetailed}.  This shows that, as long as the Higgs results are to be trusted, the \ww cross section anomaly will persist when ATLAS and CMS finally release their full run I \ww cross section measurements.

Ref.~\cite{Curtin:2012nn} proposed one possible explanation for this anomaly. It was shown that certain Electroweakinos could improve the $\chi^2$ of the \ww differential distributions significantly compared to the SM, while evading all other direct searches at the time. Subsequent to this, it was shown that scenarios involving a single squeezed stop~\cite{Rolbiecki:2013fia} or light sleptons \cite{Curtin:2013gta} could also fit the data.  In this paper we show that there are several more scenarios involving stops than the one proposed in~\cite{Rolbiecki:2013fia} that can also fit the \ww anomaly.  In particular we show that there are scenarios where the third generation alone plays the role of generating the signal, rather than relying upon a particular squeezing between a stop and chargino as in~\cite{Rolbiecki:2013fia}.  Additionally, we also show that {\em both} stop eigenstates can be light and explain the \ww signal, thereby satisfying all naturalness requirements in the most important sector of SUSY models. Finally it is also possible, in principle, to combine these results with  previous findings in~\cite{Curtin:2013gta}, where the $(g-2)_\mu$ anomaly and the relic density of DM in the universe are also explained.

In considering these light stop scenarios we do not address the Higgs mass within SUSY, implicitly relying on one of the above-mentioned mechanisms for generating additional contributions needed to account for the observed value of $\approx 125 \gev$.  This puts the discussion of naturalness within SUSY on equal footing with, for instance, many composite Higgs models~\cite{comphiggsreview}.  In principle the spectra and types of particles investigated here do not have to be realized within a supersymmetric framework, and an alternative model with top partners could also explain the \ww excess with low mass particles. 

 Even putting aside the Higgs mass, there are other measurements that can indirectly bound stops by their radiative contributions to Higgs couplings~\cite{ewbghiggs,Fan:2014txa ,Carena:2013iba}. The introduction of light stop partners can significantly enhance the $h\rightarrow gg$ production process, constraining the mass scales we are interested in. However, these constraints rely on combined coupling fits, and the differences between ATLAS and CMS measurements significantly weaken constraints~\cite{Fan:2014txa, mattemail}. 

Taking all this into account, along with other relevant bounds from direct searches, we demonstrate that stops can still be very light, allowing them to contribute their part of the naturalness puzzle while simultaneously fitting the LHC data better than the SM alone. 

This paper is organized as follows. In \sref{bsmexplanations} we review previously suggested BSM explanations for the \ww excess, and summarize the main features of the light stop scenarios we study in this work. \sref{scenarios} defines each scenario and studies its phenomenology in detail, discussing improved fit to the data in the \ww measurement,  potential  simultaneous explanation for the DM relic density and the anomalous $(g-2)_\mu$, and bounds from direct searches and Higgs couplings. We conclude in \sref{conclusion}, with some technical details of the Monte Carlo simulations outlined in \aref{montecarlo}.

\section{BSM Explanations for the \ww Excess}
\label{s.bsmexplanations} \setcounter{equation}{0} \setcounter{footnote}{0}

The BSM scenarios in~\cite{Curtin:2012nn} and~\cite{Curtin:2013gta} explained the observed \ww excess using electroweak production of new particles, while \cite{Rolbiecki:2013fia} utilized strong production channels. In each case, the new particles decay to a $\ell \ell + \mathrm{MET}$ observable final state and mimick the dileptonic \ww signal. Any such spectrum has to escape detection by a multitude of new physics searches for lepton-rich final states. Ultimately this led to a handful of viable scenarios to explain the \ww excess while remaining consistent with all other LHC data, which we review briefly below. We also outline the new light stop scenarios we study in this work. 

In \cite{Curtin:2012nn} we explored electroweak production of charginos decaying into $W + \mathrm{LSP}$. At a mass of $\sim 110 \gev$, a wino-like chargino has the required direct production cross section of a few pb to explain the \ww excess. However, this possibility is ruled out in simple gravity-mediated scenarios, since $\chi^0_2 \chi^\pm_1$ associated production yields a large $WZ$ signal which is thoroughly excluded at that mass scale \cite{atlastrilepton, cmstrilepton}. While Higgsino-like scenarios above the LEP limit are not yet excluded \cite{CMS:2013afa, TheATLAScollaboration:2013zia}, their chargino pair production cross section is too small to explain the \ww excess. This led us in \cite{Curtin:2012nn} to consider a gauge-mediated scenario \cite{Kribs:2008hq} with a $\approx 110 \gev$ chargino NLSP decaying to a massless gravitino. Neutralinos $\chi^0_{1,2}$ at $\approx 113$ and $130 \gev$ decay to charginos via off-shell $W^\pm$ emission, which is mostly too soft to be detected. This further enhances the chargino signal. Adding the chargino contribution to the \ww signal expectation in \cite{ATLAS:2012mec, CMS:2012tbb, CMS:2012daa} greatly improves fit to data, both in terms of overall cross section and shape agreement in all differential distributions. Strikingly, the signal bins in which the SM correctly accounts for the data are not modifed, while the chargino contribution is concentrated in exactly those bins where the SM expectation is below the data. A side-effect of this spectrum is a sizable same-sign dilepton signature, which serves as a smoking gun of the chargino NLSP scenario.

The \ww excess could also be explained without producing any actual $W$-bosons. In \cite{Curtin:2013gta} we showed that $\sim 130 \gev$ sleptons decaying to dileptons and $\sim 75 \gev$ Binos also have the correct cross section and kinematics to account for the \ww anomaly.\footnote{$\tilde \ell \to W \tilde \nu$ is not suitable. $m_{\tilde \ell_L} - m_{\tilde \nu_L}$  is too small for LH slepton production to give correct kinematics for the $\ell \ell$ + MET final state, but large enough for it to be excluded by LEP searches if the RH slepton is on top of the spectrum to explain the \ww excess.} The light slepton scenario is compelling, since the spectrum preferred by \ww also generates the correct dark matter relic density by providing a sufficiently large $t$-channel annihilation process for the Bino, and explains the anomalous $(g-2)_\mu$ measurement. The smoking gun of this possibility is a predicted flavor-diagonal excess in \ww. Ref. \cite{Curtin:2013gta} also sets new constraints on slepton scenarios by using the \ww measurement as a new physics search. The observation that diboson measurements can provide new BSM constraints orthogonal to traditional high-MET SUSY searches (which cut away diboson background) is a general one, and should apply to other scenarios as well.\footnote{We checked whether the \ww measurements provide new constraints on chargino pair production scenarios, but the low cross section and \emph{preference} of \ww data for light charginos means that in this case no new constraints can be derived. Ref. \cite{ATLAScharginotoWMETdileptonsearch} directly searched for $\tilde \chi^\pm_1 \to W + \tilde \chi^0_1$, and also specifically for the Chargino model presented in \cite{Curtin:2012nn}, but does not have sensitivity to cross sections relevant for SUSY.}

The above two possibilities involve relatively simple spectra, but the scale of new physics has to be lower than about $150 \gev$, otherwise the electroweak production cross sections are too low to account for the \ww excess. This restriction can be avoided if the BSM states decaying to $W$'s (or dileptons + MET) are colored.  As mentioned in Section~\ref{s.intro}, \cite{Rolbiecki:2013fia} proposed a squeezed stop scenario where a relatively light stop decays to a chargino (and a soft, presumed undetectable $b$) with a mass gap of $m_{\tilde t_1} - m_{\tilde \chi^\pm_1} \lesssim 10 \gev$. In \fref{scenarios} this is called Scenario A. It effectively gives the chargino a strong production cross section, allowing it to be as heavy as $\sim 250 \gev$ while still providing enough events in the \ww signal region to potentially explain the excess. The authors of \cite{Rolbiecki:2013fia} performed no differential analysis within the signal region, but to replicate the kinematic shape fit of our original chargino scenario \cite{Curtin:2012nn}, the mass difference between the chargino and neutralino LSP would have to be about $m_W$. We will confirm this in the next section.

In this work we suggest a qualitatively different mechanism for accounting for the \ww excess via QCD production as well as two other extended scenarios. Rather than using stops to produce electroweakinos, $W$'s can be produced directly from electroweak stop decay to a light sbottom, which then has to be close in mass to a neutralino LSP to be undetectable. This is Scenario B in \fref{scenarios}. In \sref{scenarios} we perform a fully differential fit of both single stop scenarios to the \ww data, identifying the regions in the stop-neutralino mass-plane that are preferred (or excluded) by the \ww measurement while escaping stop and sbottom direct search constraints. The best-fit point for both single stop scenarios is near $(m_{\tilde t_1}, m_{\tilde \chi^0_1}) \sim (220, 130) \gev$.

While Scenarios A and B provide intriguing explanations of the \ww excess using colored particles, ultimately light stops are theoretically motivated for reasons of naturalness. 
The single light stop scenarios are certainly interesting in this regard, but in both cases the rest of the third generation squarks has to generically be heavy (near a TeV) to avoid direct stop and sbottom searches \cite{stoptotopsearches, ATLASsb2}. Therefore, in those cases naturalness in the stop sector is only partially accommodated.  This motivates us to explore the possibility of not just one stop, but both stops and at least one sbottom below $\sim 250 \gev$, shown in \fref{scenarios} as Scenarios C and D.  In both cases the stops are close in mass and decay either to charginos or to $W$'s directly, generalizing the above single-stop scenarios. As we will see below, both of these scenarios are viable, meaning the \ww excess could already be pointing towards a completely natural light SUSY spectrum.  
There are of course indirect constraints on light third generation sectors. For instance, split LH squarks are subject to EW oblique constraints~\cite{obliquesusy}. Natural theories with light charginos and third generation squarks can also generate deviations to $b\rightarrow s\gamma$, as most recently shown in~\cite{bsgamma}. 
These constraints are easily accommodated in scenarios A and B by making the light squarks mostly RH. In scenarios C and D a careful analysis of the indirect constraints is a priori necessary. We do not pursue this line of enquiry here, since the unspecified additional sectors, which account for the observed higgs mass, could also reduce any loop-generated indirect signatures of a light third generation.

Going beyond \ww and naturalness, the new stop scenarios we propose could also replicate some of the phenomenological success of the slepton scenarios in \cite{Curtin:2013gta}. Firstly, the presence of light sbottoms could make the Bino DM a thermal relic. Secondly, in the absence of a chargino (Scenarios B and D), sleptons could sit between the LSP and the stop(s). The light smuon could then account for the $(g-2)_\mu$ anomaly without being excluded by direct searches. A plethora of new particles may await discovery below 250 GeV.

\begin{figure}
\begin{center}
\hspace*{-7mm}
\includegraphics[width=18cm]{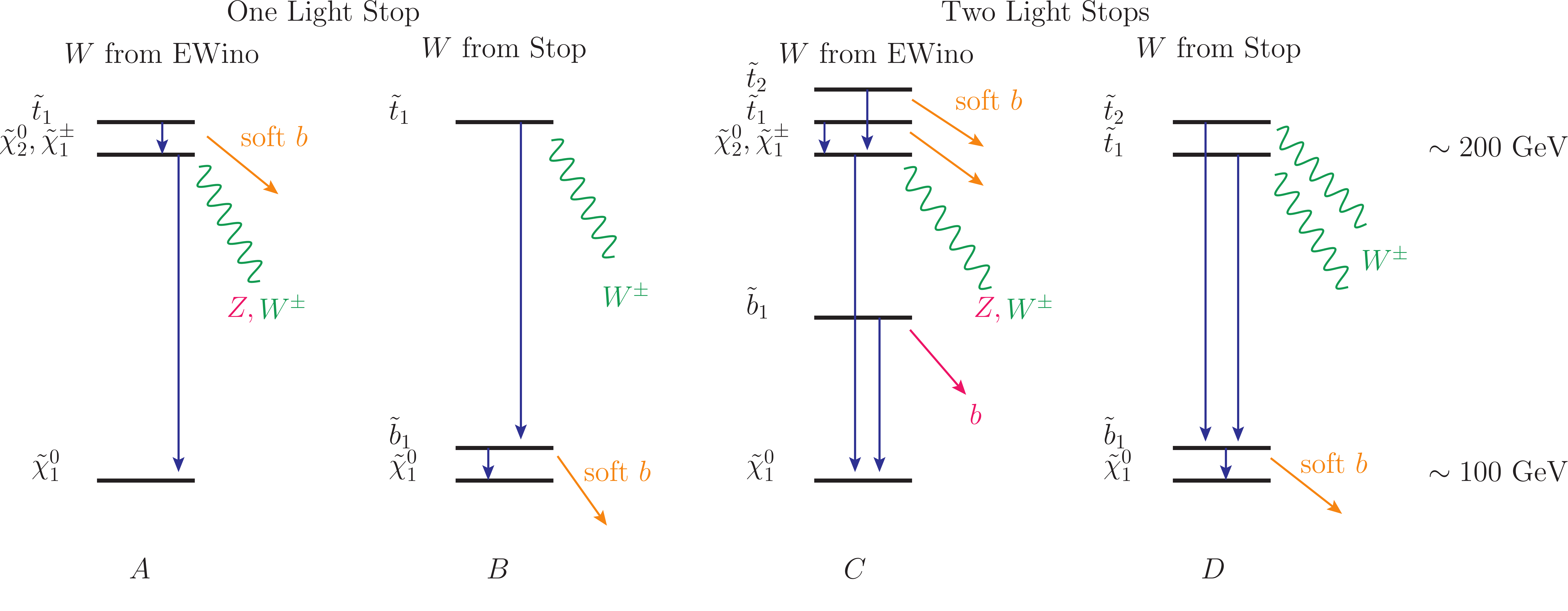}
\end{center}
\caption{
The four types of stop spectra which could account for the \ww excess via stop pair production, labelled Scenarios A - D. The top and bottom of the spectrum are at $\sim 200 \gev$ and $\sim 100 \gev$, with $W$'s (green) being produced when decaying across the big gap in the spectrum. Small gaps are $\lesssim 10 \gev$. The 2-body decays of each state are shown as blue vertical arrows, with SM decay products on the right of each spectrum. The red color for $Z$ and $b$ indicates that these are not produced from stop pair production but from a different processes (direct $\tilde \chi^0_2 \tilde \chi^\pm_1$ and $\tilde b_1 \tilde b_1^*$ production). The soft $b$'s (orange) should be practically undetectable.  
}
\label{f.scenarios}
\end{figure}

\section{Light Stop Scenarios}
\label{s.scenarios} \setcounter{equation}{0} \setcounter{footnote}{0}

In this section we will show how each of the light stop scenarios in \fref{scenarios} could account for the \ww excess. In each case a $\chi^2$-fit over all kinematic distributions of the \ww cross section measurements \cite{ATLAS:2012mec, CMS:2012tbb, CMS:2012daa} is performed, with preferred regions of the stop-neutralino mass-plane identified by smaller values of $\chi^2_\mathrm{SM+stops}/\chi^2_\mathrm{SM}$. Details of the fit and Monte Carlo simulation are included in \aref{montecarlo}.  We include in our analysis the constraints from stop, sbottom and chargino direct searches, and find that they do not exclude one or two light stops as explanations for the \ww excess. In fact, as we outline below, chargino searches may already hint at an independent confirmation of certain types of spectra.   

The presence of light sbottoms in Scenarios B-D allows the Bino to be a thermal DM candidate with correct relic density. The absence of charginos in Scenarios B \& D also allows light sleptons to be included, which can account for the measured deviation in $(g-2)_\mu$. The corresponding treatment of these issues for Scenario B in \sssref{dmgm2} carries over to the subsequent scenarios. We also discuss Higgs coupling constraints on Scenario C \& D with two light stops in \sssref{higgscoupling}. They are not prohibitive, but will be an interesting probe at the next run of the LHC.

\subsection{Scenario A: One Light Stop, $W$ from EWino}
\label{ss.scenarioA}

\begin{figure}[htbp] 
\vspace*{-1.5cm}
\begin{center}
\hspace*{-1.5cm}
 \begin{tabular}{cc}
 \includegraphics[width=8.6cm]{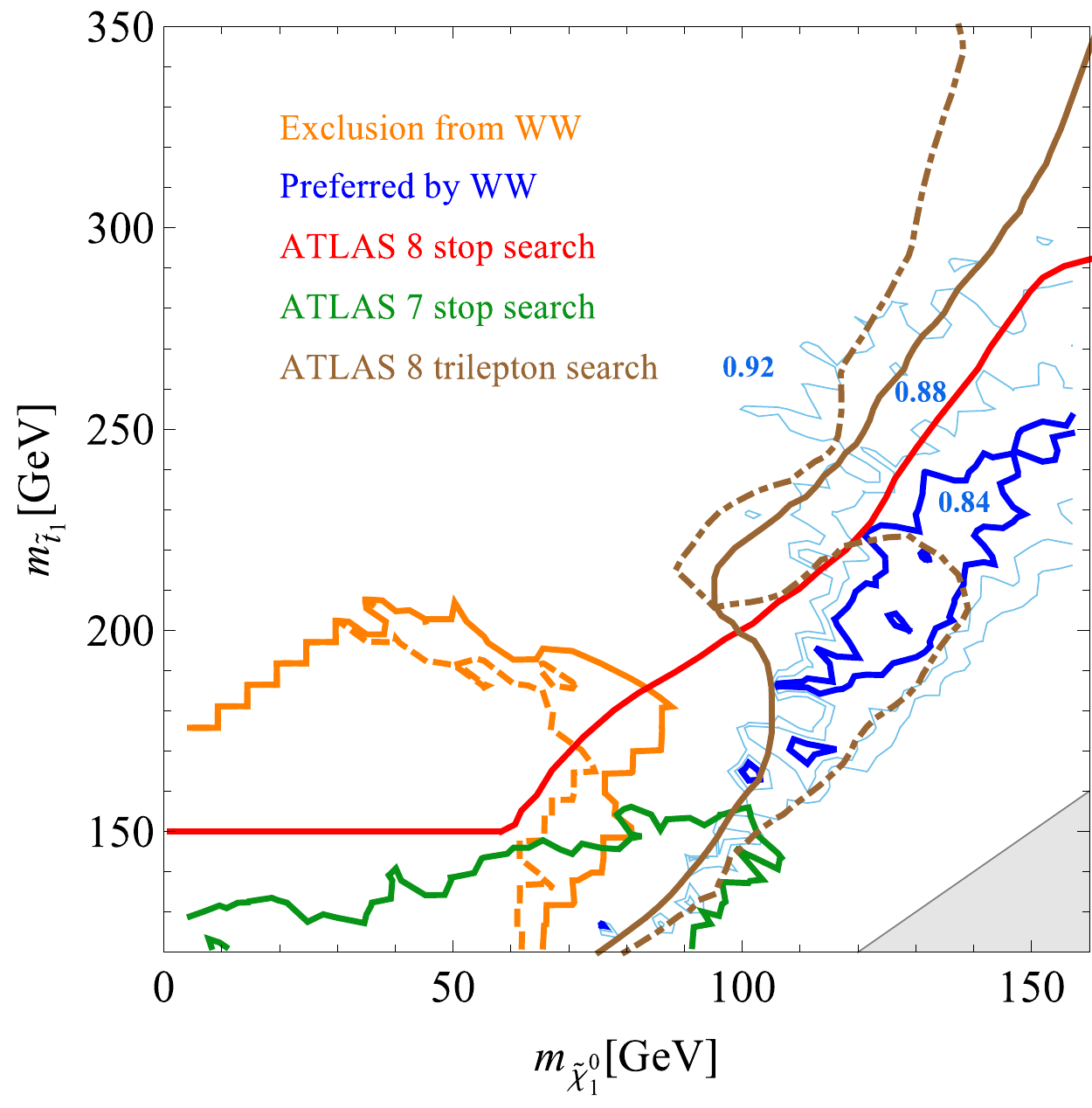} & 
 \includegraphics[width=8.6cm]{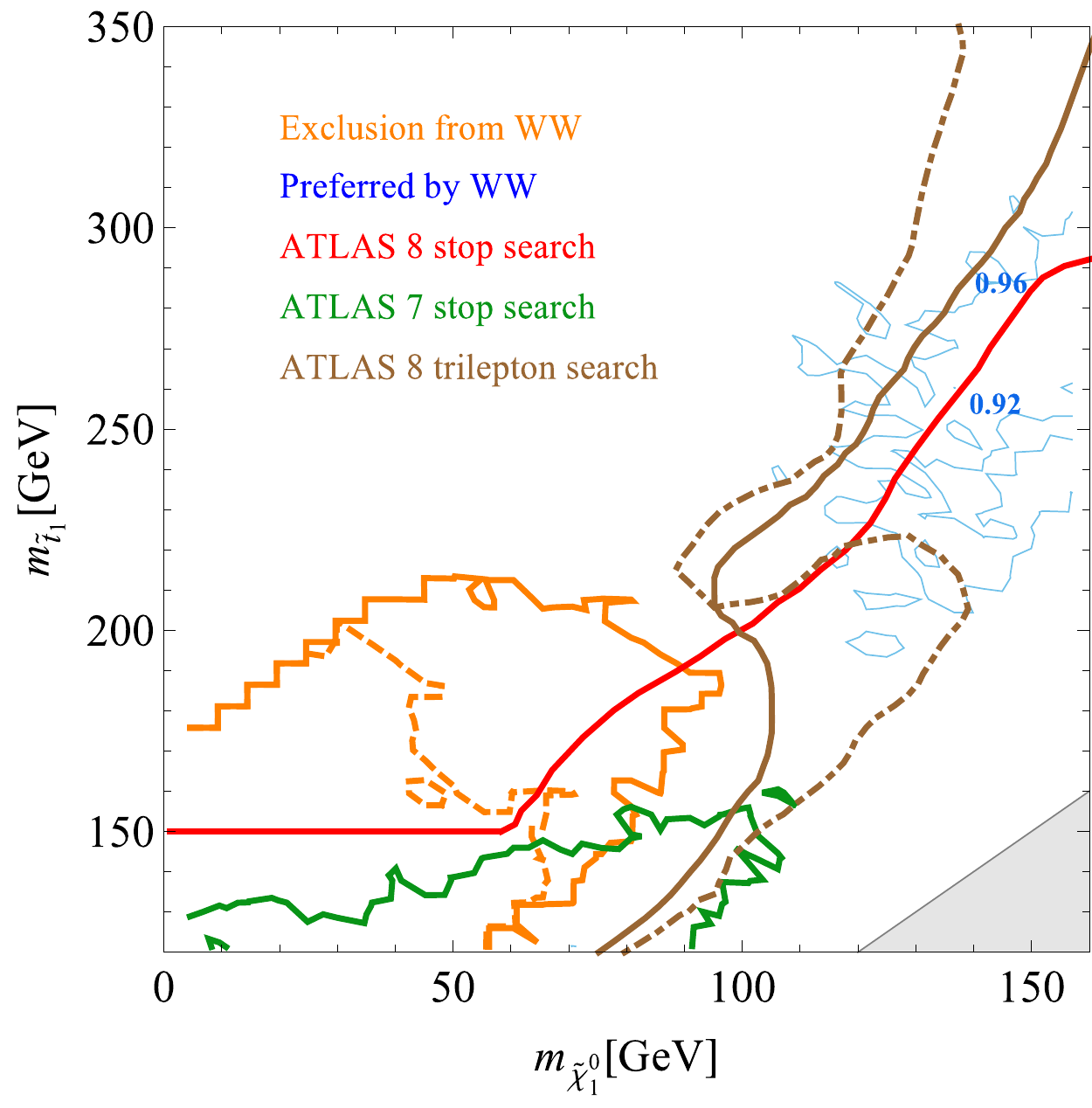} \\ (a) ATLAS 7 TeV 5 $\ifb$ \cite{ATLAS:2012mec} & (b) CMS 7 TeV 5 $\ifb$ \cite{CMS:2012tbb} \vspace{2mm}
 \end{tabular}
 \hspace*{-1.5cm}
 \begin{tabular}{cc}
  \includegraphics[width=8.6cm]{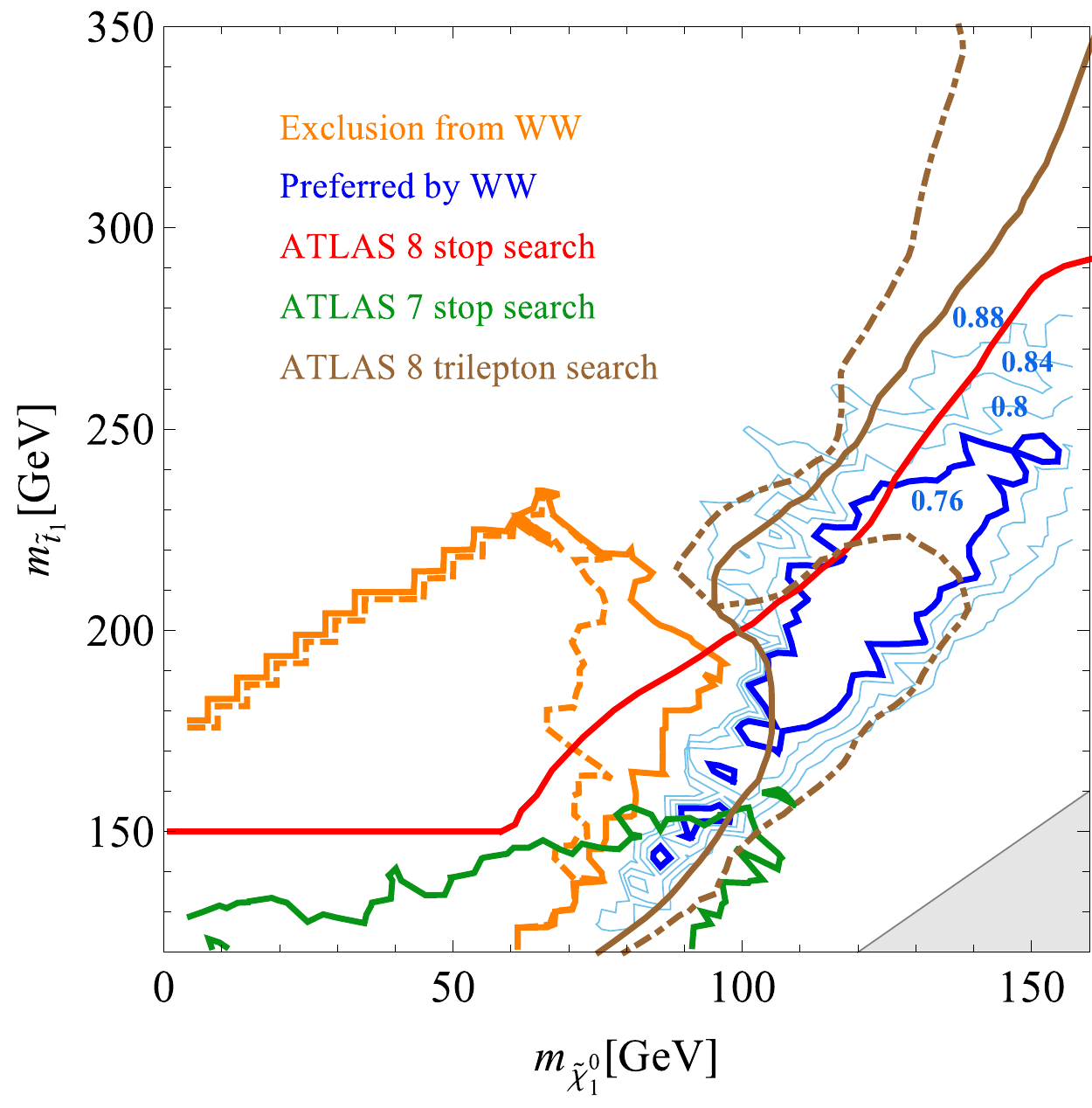}  \\ (c) CMS 8 TeV 3.5 $\ifb$ \cite{CMS:2012daa}
 \end{tabular} 
    \caption{
   Regions of the stop-neutralino mass-plane excluded and preferred by the different \ww cross section measurements in Scenario A ("One Light Stop, $W$ from EWino"). We fix $\Delta m = \tilde{t}_1 - \chi^\pm_1 \approx 10\gev$ to avoid hard b-jets. 
Solid (dashed) orange line: $95\%$ exclusion from the  \ww measurement with fixed (floating) normalization of SM contribution.
Thin blue contours show values of $\chi^2_\mathrm{SM+stops}/\chi^2_\mathrm{SM}$, with the thick contour indicating the region most preferred by the \ww measurement. 
Exclusions from ATLAS stop searches shown in red \cite{ATLAS8stop1} and green \cite{ATLAS7stop1}. Observed (expected) exclusion from ATLAS trilepton $\chi^0_2 \chi^\pm_1$ search \cite{atlastrilepton} shown as solid (dot-dashed) brown line: note how an excess compatible with the \ww preferred region pushes the observed bounds down in Bino mass.
}
\label{f.stopww1}
\end{center}
\end{figure}

This is Scenario A in \fref{scenarios}, originally proposed by \cite{Rolbiecki:2013fia}. A single light stop is pair-produced and decays via soft $b$-jets to wino-like charginos, which then decay to a $W$ and a Bino LSP. The second stop could evade detection if it hides in the $t \bar t$ background with a mass of $m_{\tilde t_2} \approx m_t + m_{\tilde \chi^0_2}$, but then sbottom constraints would exclude this scenario, see \fref{sbottomexclusion}. Therefore we assume the second stop to be heavier than $\sim 700 \gev$ to evade $\bar t t + \mathrm{MET}$ searches \cite{stoptotopsearches}.

\fref{stopww1} shows the stop-neutralino mass plane, with $m_{\tilde \chi^\pm_1} \approx m_{\tilde t_1} - 10 \gev$. (If the mass difference were much larger the stop events would fail the jet veto of the \ww measurements.) The region \emph{above} the red contour is excluded by the 13 $\ifb$ ATLAS  8 TeV low-MET $\tilde t \to b + \tilde \chi^\pm_1$ search.\footnote{A recent 20 $\ifb$ update \cite{Aad:2014qaa} does not significantly change the limits in our mass region of interest.} Lighter stop masses $m_{\tilde t_1} < 150 \gev$ are constrained by a 5 $\ifb$ 7 TeV ATLAS search \cite{ATLAS7stop1}. Applying the cuts from this search, and rescaling our efficiency by 0.5 to reproduce the acceptances quoted in \cite{ATLAS7stop1}, excludes the region below the green curve. Finally, the observed (expected) limits on $\chi^0_2 \chi^\pm_1 \to W + Z + 2 \chi^0_1$ from the ATLAS 20$\ifb$ 8 TeV trilepton search \cite{atlastrilepton} are shown as a solid (dot-dashed) brown line. Note the deviation between observed and expected chargino limits, which is due to a $2 \sigma$ excess in the SR0$\tau$a-bin01 of that search.  

The solid (dashed) orange line shows the constraint obtained on this stop scenario by each of the published \ww measurements under the assumption of fixed (freely floating) SM contribution. The obtained limits close the gap between the two stop searches, but are superseded (in this scenario) by the trilepton limits. 

The thin blue lines are contours of $\chi^2_\mathrm{SM + stop}/\chi^2_{SM}$ for the full shape fit across all published differential distributions in each \ww search. The actual value of this ratio is not very meaningful, since the public data does not allow us to take all correlations into account for the shape fit. Nevertheless, the result that some regions in the mass plane are preferred over others and improve the fit compared to the SM alone is robust, and we indicate the ``most preferred regions'' with a thick blue contour to guide the eye. Its vertical extent is mostly given by the stop production cross section. A stop-neutralino mass-difference of $\sim m_W$ is preferred to give roughly at-rest $W$'s from chargino decay, improving agreement in all kinematic distributions of the \ww measurements. (If the kinematics were very different, the stop contribution would fill in the wrong bins and worsen the disagreement between expectation and data.) The best-fit point is near $(m_{\tilde t_1}, m_{\tilde \chi^0_1}) \approx (220, 130) \gev$.

The $WW$-preferred region is not excluded by either stop or chargino bounds. In fact, the ATLAS trilepton search \cite{atlastrilepton} should be sensitive to the stop spectra in part of the preferred region, but the observed $2 \sigma$ excess pushes the exclusion away from the preferred region. This might be interpreted as very tentative evidence for this light stop scenario, from a signal which is completely uncorrelated with the dilepton + MET final state in the \ww measurement.\footnote{The CMS trilepton search \cite{cmstrilepton} has no sensitivity in this mass region.}

The pure Bino is a slightly problematic DM candidate within the MSSM, requiring non-standard cosmological history to have the correct relic density. This is discussed further in \sssref{dmgm2}.

\subsection{Scenario B: One Light Stop, $W$ from Stop}
\label{ss.scenarioB}

In contrast to the first example where charginos were required to produce the $W's$ in their decays, $W's$ can be produced with colored cross section simply via electroweak stop decay. This is Scenario B in \fref{scenarios}.

$\tilde t_2$ is again assumed to be heavier than $\sim 700 \gev$ to evade direct searches and demonstrate the minimal working parts necessary. The presence of a light sbottom decaying via $\tilde b_1 \to b + \tilde \chi^0_1$ is highly constrained, most importantly by a a 12.8 $\ifb$ ATLAS search \cite{ATLASsb2}, see \fref{sbottomexclusion}. However, these bounds can be avoided if $m_{\tilde b_1} - m_{\chi^0_1} \lesssim 10 \gev$, since for such small mass gaps sbottom decay is poorly understood, and it is possible for such spectra to evade searches by failing $b$-jet requirements or single-track vetoes.

\begin{figure}
\begin{center}
  \includegraphics[width=9cm]{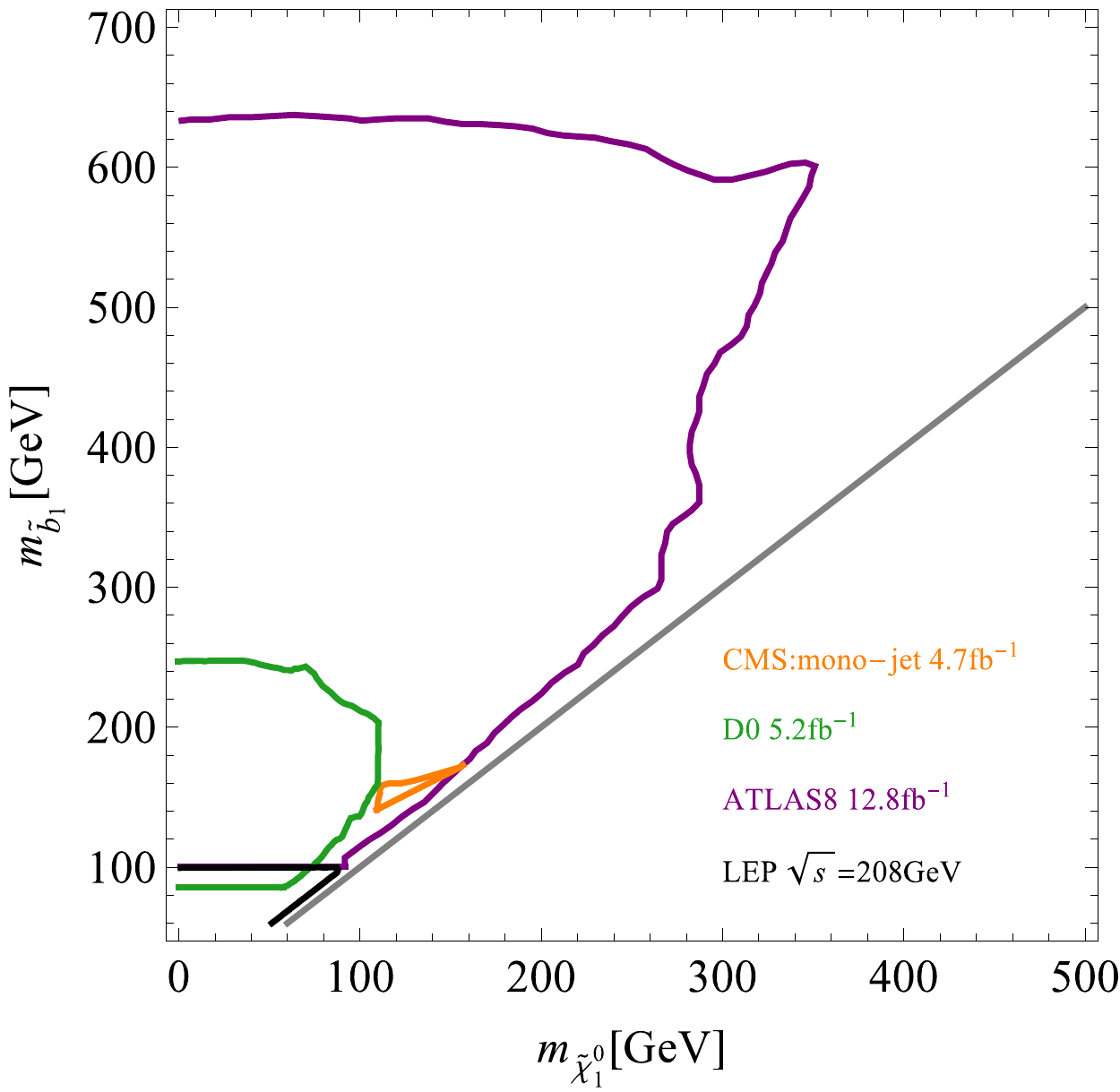}
    \caption{Bounds on a single sbottom decaying via $\tilde b_1 \to b + \tilde \chi^0_1$. Black: LEP $\sqrt{s}=208\gev$ ~\cite{LEPsb}. Purple: low-MET ATLAS 8TeV 12.8 $\ifb$  search \cite{ATLASsb2}. Green: D0 5.2 $\ifb$ ~\cite{D0sb}. Orange: CMS 4.7 $\ifb$ mono-jet recast by \cite{Yangbai}. Gray: $m_{\tilde b_1} = m_{\tilde \chi^0_1}$ kinematic limit.}
\label{f.sbottomexclusion}
\end{center}
\end{figure}

Again for simplicity we assume mostly right-handed $\tilde t_1$ and $\tilde b_1$ to decouple $m_{\tilde b_1}$ from $m_{\tilde t_1}$ and easily allow for  $m_{\tilde b_1} \sim m_{\chi^0_1}$. (Mixed sbottoms can also be accommodated by adjusting sbottom mixing, see \ssref{scenarioC}.)
Both states, $\tilde t_1$ and $\tilde b_1$, have to carry at least a small LH component to ensure $\mathrm{Br}(\tilde t_1 \to \tilde b_1 + W^+) \approx 1$ and avoid a large $\tilde t \to c + \tilde \chi^0_1$ signal. Higgs coupling measurements are not yet sensitive to a single light stop \cite{Fan:2014txa}, while deviations due to sbottoms are generically small,  certainly so if the other sbottom is very heavy \cite{Blum:2012ii}.

The kinematics of the BSM signal in the \ww measurement is very similar to Scenario A, so most of \fref{stopww1} applies here as well. The same stop search limits apply, but there are no bounds from the ATLAS trilepton searches since there is no light wino-like chargino/neutralino pair.  With $m_{\tilde b_1} - m_{\chi^0_1} \approx 10 \gev$ there are no sbottom bounds, and a nearly identical region of the stop-neutralino mass plane is preferred/excluded by the \ww measurements. In the absence of a trilepton signal, these new bounds fill an important gap between the stop searches.

\subsubsection{Thermal Bino Dark Matter and $(g-2)_\mu$}
\label{sss.dmgm2}

The pure Bino is a slightly problematic dark matter candidate within the MSSM. If it is the LSP, its annihilation cross section is typically very small, leading it to overclose the universe. (For a discussion see e.g. \cite{welltempered}.) Scenario A can therefore not be realized within the standard MSSM, and some additional mechanisms to dilute the Bino density must be present. 

Bino annihilation can be enhanced in three ways. Firstly, if the Bino-like LSP has a non-negligible  Wino (Higgsino) fraction and its mass is near $m_Z/2$ ($m_h/2$),  annihilation  proceeds through an $s$-channel $Z$ ($h$) resonance. Secondly, if there is another sfermion close in mass it is possible to co-annihilate both LSP and NLSP particle populations. Thirdly, if there is a relatively light sfermion carrying hypercharge then it can mediate sizable annihilation via $t$-channel exchange. 
Scenarios B - D feature light sbottoms between the LSP and stops in the spectrum. The presence of this additional degree of freedom makes it possible to enhance Bino annihilation to either make it a subdominant dark matter component, or to act as a thermal relic  with the correct relic density $\Omega_{\mathrm{CDM}} h^2 = 0.1196 \pm 0.0031$ \cite{planck}. 

To understand the impact of a light sbottom we computed the Bino DM relic density $\Omega_\mathrm{Bino}$ using \texttt{micrOMEGAs 3.6.9.2} \cite{micromegas} for different $m_{\tilde \chi^0_1}, m_{\tilde b_1}$ assuming either $\tilde b_1 = \tilde b_R$ or $\tilde b_1 = \tilde b_L$.\footnote{We assume $m_h = 125 \gev$ is generated by the heavy second stop or by some new physics beyond the MSSM for the scenarios with two light stops, so we fix the Higgs mass manually in the SLHA spectrum files.} In either case, we find that $t$-channel annihilation is insufficient to avoid overclosure, due to the small hypercharge of sbottoms compared to sleptons. The only way to satisfy $\Omega_\mathrm{Bino} = \Omega_\mathrm{CDM}$ with light sbottoms is via co-annihilation. For the Bino masses most of interest, $m_{\tilde \chi^0_1} \sim 130 \gev$, this requires $m_{\tilde b_1} \approx m_{\tilde \chi^0_1} + 15 \gev$ for both $\tilde b_L$ and $\tilde b_R$. This is just on the border of exclusion in the ATLAS sbottom search \cite{ATLASsb2} (see \fref{sbottomexclusion}), so this mechanism for generating the correct thermal relic density may be called marginally viable. At any rate, if the sbottom is closer in mass to the Bino than 15 GeV then the Bino makes up some fraction of the total DM density. This means the light sbottom scenarios are \emph{not} excluded by cosmological considerations. 

Regardless of cosmological history, if a Bino-like LSP constitutes a significant dark matter component then its higgsino fraction must be low enough to give a Higgs-mediated direct detection cross section below current bounds. We checked that LUX direct detection bounds \cite{luxdm} are satisfied for $\mu \gtrsim 500 \gev$. 

Sbottom-Bino co-annihilation can make the LSP in Scenarios B - D a thermal relic in the $WW$-preferred region. There is, however, potential to address yet another  anomaly which may hint at new physics. The absence of charginos in Scenarios B and D makes it possible to insert sleptons into the spectrum between the stop and the LSP without affecting the \ww signal from stop pair production. High-MET SUSY searches are not sensitive to sleptons in the ``$WW$-funnel'', $m_{\tilde \ell} - m_{\tilde \chi^0_1} \lesssim m_W$ \cite{atlastrilepton}. In \cite{Curtin:2013gta} we showed that such sleptons below $\sim 150 \gev$ could account for the \ww anomaly while simultaneously providing a thermal Bino relic and serving as an explanation for the long-standing $3\sigma$ deviation in the measured value of the muon anomalous magnetic moment $(g-2)_\mu$ \cite{pdg:2012}. Inserting sleptons above $\sim 150 \gev$ into the spectrum of Scenarios B and D would not significantly affect the \ww signal or the relic density (which is annihilated away by sbottom co-annihilation) but the light smuon could still explain $(g-2)_\mu$.

In summary, light stop Scenario B can explain the \ww excess, while also generating the correct thermal Bino relic density and accounting for the venerable $(g-2)_\mu$ anomaly.

The conclusions of this subsection regarding relic density and direct detection can be applied verbatim to the next two scenarios as well, since they do not meaningfully depend on the stop spectrum or the composition of the lightest sbottom quark.

\subsection{Scenario C: Two Light Stops, $W$ from EWino}
\label{ss.scenarioC}

In the context of naturalness, one light stop is good but two light stops are better.   In this section and the next we will demonstrate that Scenarios A and B can be modified to have two light stops.

Scenario C in \fref{scenarios} represents a simple extension on Scenario A, making the second stop similarly light as the first one. The mass difference between the two stops has to be fairly small to ensure that $b$-jet from $\tilde t_2 \to \tilde \chi^\pm_1 + b$ decay does not trigger the jet veto in the \ww measurements. This means the stops cannot have large mixing. 

Making both unmixed stops near-degenerate will also introduce the left-handed sbottom into the spectrum. Using the notations of \cite{djouadi2}, setting stop mixing to zero $(X_t = 0)$ via judicious choice of $A_t$ for a given $\mu$ and $\tan \beta$ fixes the left-handed 3$^\mathrm{rd}$ generation squark soft mass at tree-level to be
\begin{equation}
M_Q^2 = m_{t_2}^2 - m_t^2 + \frac{1}{6} M_Z^2(4 \sin^2 \theta_W - 3) \cos 2\beta,
\end{equation}
where we take $m_{t_2}$ to be the LH stop mass. (In practice there will also be some small stop mixing and hence mass difference, to ensure both stops can decay to a chargino.) For zero sbottom mixing, this gives a LH sbottom mass
\begin{eqnarray}
\nonumber m_{b_L} &=& \sqrt{
m_{t_2}^2 + m_b^2 - m_t^2 + M_Z^2 (\sin^2 \theta_W - 1)\cos 2 \beta}
\\
\label{e.msbottom}
&\approx& 1.6 m_{t_2} - (200 \gev),
\end{eqnarray}
where the approximation in the second line holds to a few GeV in our stop mass range of interest  $m_{\tilde t} \sim 180 - 260 \gev$ when $\tan \beta \gtrsim 3$. Without sbottom mixing we therefore expect most of this Scenario's parameter space to be ruled out by sbottom searches. However, one can always lower the mass of the lightest sbottom by increasing mixing to satisfy $m_{\tilde b_1} - m_{\tilde \chi^0_1} \lesssim 10 \gev$, which removes sbottom constraints as discussed for Scenario B in \ssref{scenarioB}. The presence of light sbottoms could also help generate a thermal Bino DM relic (or annihilate away the primordeal Bino abundance so it is a subdominant dark matter component), see \sssref{dmgm2}.

Two stops near 200 GeV would make the SUSY spectrum very natural, but within the MSSM they can not generate sufficient loop corrections to lift the Higgs mass to 125 GeV. There are, however, a myriad of extensions to the MSSM which introduce additional Higgs mass contributions. As outlined in Section \ref{s.intro} we will therefore assume some such contribution is present, and concentrate on direct consequences of these light stops. 

\fref{stopww2} shows the stop-neutralino mass plane for this scenario with $m_{\tilde t_2} \approx m_{\tilde t_1}$ and small sbottom mixing. The labeling is the same as \fref{stopww1}, and the region preferred by each \ww measurement is shown by the thick blue contour. The purple line indicates the constraint from the ATLAS sbottom search \cite{ATLASsb2}. For unmixed sbottoms it excludes much of the  $WW$-preferred region, though some remains. However, increasing sbottom mixing can remove the this constraint. The fully natural scenario with \ww from electroweakinos is therefore viable, and the trilepton excess in \cite{atlastrilepton} could still be taken as tentative corroboration of this spectrum.

\begin{figure}[htbp] 
\vspace*{-1.5cm}
\begin{center}
\hspace*{-1.5cm}
 \begin{tabular}{cc}
 \includegraphics[width=8.6cm]{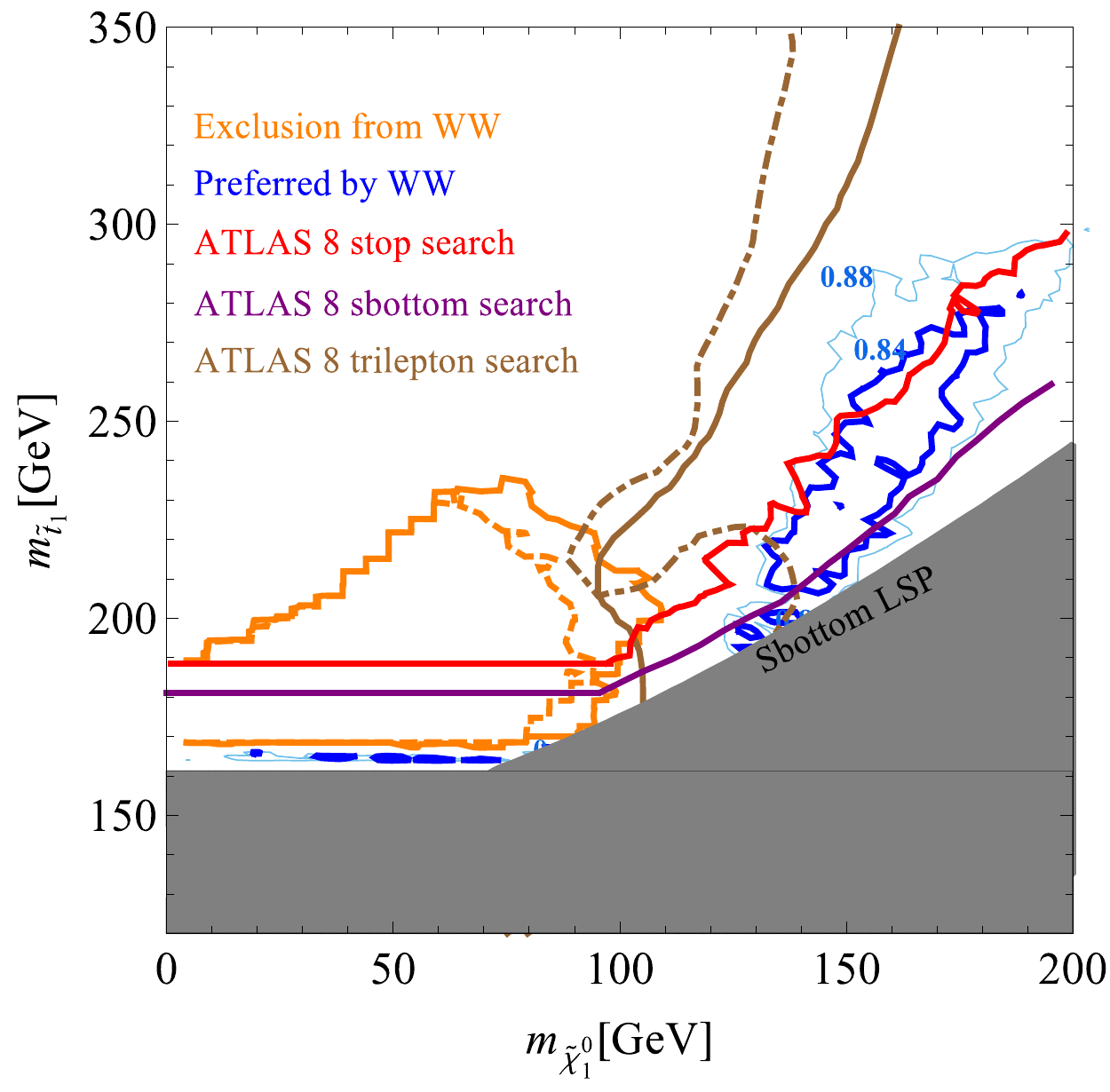} & 
 \includegraphics[width=8.6cm]{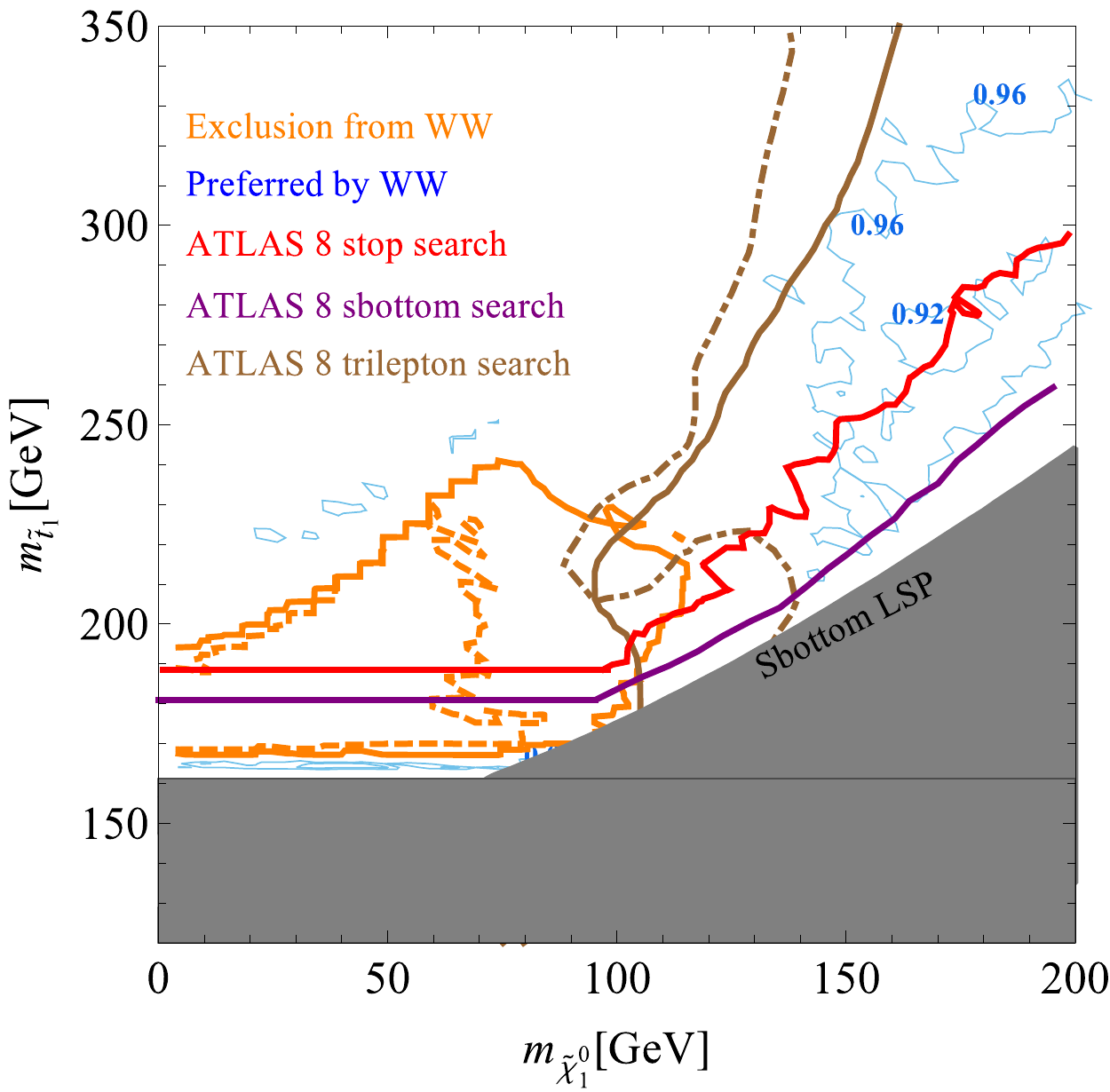} \\ (a) ATLAS 7 TeV 5 $\ifb$ \cite{ATLAS:2012mec} & (b) CMS 7 TeV 5 $\ifb$ \cite{CMS:2012tbb}  \vspace{2mm}
 \end{tabular}
 \hspace*{-1.5cm}
 \begin{tabular}{cc}
  \includegraphics[width=8.6cm]{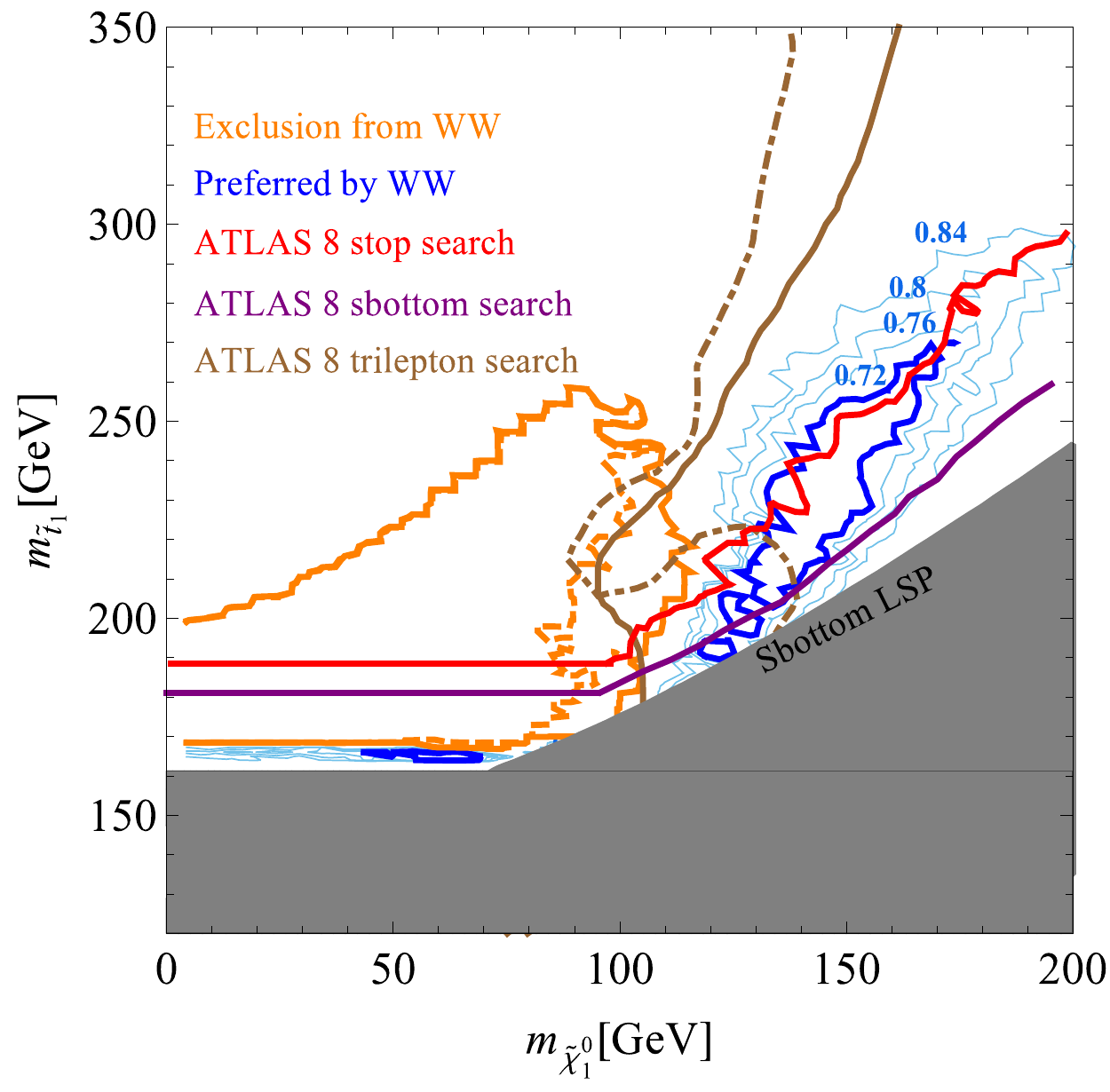}  \\ (c) CMS 8 TeV 3.5 $\ifb$ \cite{CMS:2012daa}
 \end{tabular} 
    \caption{
Regions of the stop-neutralino mass-plane excluded and preferred by the different \ww cross section measurements in Scenario C ("Two Light Stops, $W$ from EWino"). We fix $\Delta m = \tilde{t}_1 - \chi^\pm_1 \approx 10\gev$ to avoid hard b-jets, and make the two stops degenerate $m_{\tilde t_1} \approx m_{\tilde t_2}$. There is no large sbottom mixing, so $m_{\tilde b_1}$ is given by \eref{msbottom}.
Solid (dashed) orange line: $95\%$ exclusion from the  \ww measurement with fixed (floating) normalization of SM contribution.
Thin blue contours show values of $\chi^2_\mathrm{SM+stops}/\chi^2_\mathrm{SM}$, with the thick contour indicating the region most preferred by the \ww measurement. 
Exclusions from the ATLAS stop search shown in red \cite{ATLAS8stop1}. Observed (expected) exclusion from ATLAS trilepton $\chi^0_2 \chi^\pm_1$ search \cite{atlastrilepton} shown as solid (dot-dashed) brown line. The purple line is the ATLAS sbottom search \cite{ATLASsb2}, but this constraint can be removed by increasing sbottom mixing. 
}
\label{f.stopww2}
\end{center}
\end{figure}

\subsubsection{Higgs Coupling Constraints}
\label{sss.higgscoupling}

Two light stops can generate significant corrections to the loop-induced Higgs couplings (see e.g. \cite{Blum:2012ii,Fan:2014txa}). Higgs signal strength measurements in different channels can already give significant constraints on such deviations. 

As discussed recently in \cite{Fan:2014txa}, these measurements naively  exclude two light unmixed stops near 200 GeV at the $3 \sigma$ level. There are, however, important caveats to this conclusion. Firstly, \cite{Fan:2014txa} assumes no other light particles in the spectrum. The presence of other Higgs coupling modifications could loosen this constraint, especially considering that two light stops already indicate the presence of additional new physics  to raise the Higgs mass beyond the MSSM expectation. Secondly, and more importantly, the  CMS \cite{CMS:ril} measurement of $h\to\gamma\gamma$ is about $2 \sigma$ lower than ATLAS \cite{ATLAS:2013oma}, which is somewhat above the SM expectation. When only ATLAS Higgs measurements are considered, two 200 GeV unmixed stops are not excluded \cite{mattemail}. 

The general lesson here is that constraints on SUSY spectra from Higgs coupling fits must be taken with a degree of caution until disagreement between the two experiments is resolved. Once the measurements converge they can be used to test Scenarios C and D.  Ignoring small sbottom corrections, the $WW$-preferred region of this scenario in \fref{stopww2} predicts a $hgg$ and $h\gamma\gamma$ coupling that is $20 - 35\%$ larger and $\approx 10\%$ smaller than the SM, respectively. The larger $hgg$ coupling results in $h\to VV^*$ signal strengths $\sim 40 - 60\%$ larger than SM, serving as an important prediction of these natural stop scenarios in the absence of other coupling corrections.

\subsection{Scenario D: Two Light Stops, $W$ from Stop}
\label{ss.scenarioD}

Direct production of \ww from stop decay can be made fully natural in a similar fashion to \ww from EWinos. This is shown as Scenario D in \fref{scenarios}. Similar to \ssref{scenarioC}, the two stops are again near-degenerate with mixing that is small but nonzero, to allow both $\mathrm{Br}(\tilde t_{1,2} \to \tilde \chi^\pm_1 b) \approx 1$. There is some mixing in the sbottom sector to guarantee $m_{\tilde b_1} - m_{\tilde \chi^0_1} \lesssim 10 \gev$ to escape sbottom searches, but the Higgs coupling correction of this mixed $\tilde b_1$  can always be made negligible with a heavy $\tilde b_2$ \cite{Blum:2012ii}. Other Higgs coupling considerations are identical to \sssref{higgscoupling} and do not exclude this scenario.

The preferred region of the stop-neutralino (or stop-sbottom) mass plane is very similar to that shown in \fref{stopww2}, except by construction the sbottom bounds do not apply, and the absence of charginos means there are no trilepton bounds. As discussed in \sssref{dmgm2} it is possible for the Bino to be a thermal relic with correct abundance, and for sleptons inserted between the stops and the neutralino to account for the deviation in the measured $(g-2)_\mu$.

\subsection{$W$ from Sbottom}
\label{ss.wfromsbottom}

One could imagine inverting the scenarios shown in \fref{scenarios}: Producing sbottoms instead of stops, and possibly hiding stops by setting their mass very close to the neutralino LSP. However, this is either not viable or already excluded.

Scenarios A and C, with $W$ from EWino decay, cannot be inverted because the $\tilde b \to \tilde \chi^\pm_1 t$ decay is $4$-body and highly suppressed if the mass difference is small, and highly visible if it is not. 

Inverted Scenarios $B$ and $D$, with one sbottom near $\sim 200 \gev$ and one or two stops near the neutralino, could generate the required \ww signal. This requires a tuned sbottom mixing to ensure $\mathrm{Br}(\tilde b_1 \to \tilde \chi^0_1 b) \ll \mathrm{Br}(\tilde b_1 \to \tilde t_{1,2} W^-) \approx 1$, which is equivalent to tuning away the effective hypercharge of $\tilde b_1$. The light stops then decay via the loop-induced process $\tilde t \to c \tilde \chi^0_1$. However, such squeezed stops are the subject of dedicated ATLAS and CMS searches \cite{stoptocharmsearches}, which exclude $m_{\tilde t} < 250 \gev$ for arbitrarily small $m_{\tilde t} - m_{\tilde \chi^0_1}$. Since the bottom of the spectrum has to be below $\sim 150 \gev$ to generate a suitable \ww signal, this eliminates the inverted Scenarios B and D as possibilities.

\section{Conclusion}
\label{s.conclusion} \setcounter{equation}{0} \setcounter{footnote}{0}

Naturalness prompts us to expect something beyond the SM near the electroweak scale. In light of this expectation, the absence of convincing new physics signals in all searches to date might be interpreted as painting a somewhat pessimistic picture. This has led to a degree of soul-searching within the field, questioning the basic assumptions on which these expectations are built. While this is a necessary exercise, it is important to understand that the possibilities for electroweak-scale new physics are far from exhausted. 

The excess in all \ww cross section measurements \cite{ATLAS:2012mec, CMS:2012tbb, CMS:2012daa} can be interpreted as (i) a statistical fluctuation, (ii) an unexplained SM effect, or (iii) a genuine signal of new physics. 

The first possibility is, by definition, somewhat unlikely, with the combined significance of the excess being about $3 \sigma$, more if shape differences in expected and observed distributions are taken into account. 

The second possibility would require very unexpected effects from QCD NNLO corrections \cite{Dawson:2013lya}. 
If there \emph{were} additional unexpected QCD behavior, it should manifest itself in the measurement of  $ZZ$ production, but both ATLAS and CMS measure that cross section to be in perfect agreement with the SM prediction \cite{ZZmeasurement}. Furthermore, the cross section for $ZZ$ production was recently evaluated at full NNLO for the first time in~\cite{zznnlo}.
The effects compared to NLO were found to be quite small provided that the $gg\rightarrow VV$ contribution to the cross section was included separately at NLO (which it is by both ATLAS and CMS in their \ww and $ZZ$ measurements).  While this result cannot be transferred verbatim to a full NNLO \ww calculation, there is reason to believe the relevant effects should be similar in size for both of these EW processes. Finally, jet veto uncertainties are addressed in a recent $p_T$-resummation calculation, which actually indicates the excess may be \emph{bigger} than reported by the collaborations \cite{resummation}. 

The third possibility has been the subject of some enquiry, both by us in this and previous papers, and other groups. Regardless of the particular interpretation, the mere fact that the excess exists means that there is either evidence or at least possible room for new physics in the \ww measurement. 

In \cite{Curtin:2012nn} and \cite{Curtin:2013gta} we showed that charginos or sleptons could account for the \ww excess, or, depending on one's interpretation, that such low-lying spectra below 150 GeV could not be excluded and remain open as possibilities. Producing $W$'s by decaying stops to charginos was first proposed in \cite{Rolbiecki:2013fia}, realizing what we call Scenario A from \fref{scenarios}. This suggested the intriguing possibility that natural SUSY spectra might be hiding in the \ww signal, or (again) at the very least are not excluded.

We showed in this paper that, in fact, several classes of spectra featuring one or two light stops can serve as viable explanations of the \ww excess without being excluded by other searches. These new possibilities are shown in \fref{scenarios}, and their phenomenological consequences are summarized in \tref{summary}. Scenario B introduces a qualitatively novel way of producing $W$'s from strong production via direct electroweak stop decay, while Scenarios C and D make both strong \ww production mechanisms fully natural. In each of these scenarios, the \ww signal is explained by one or two light stops with masses near $\sim 220 \gev$ and a neutralino LSP near $\sim 130 \gev$. All of these scenarios predict additional particles, charginos (A, C) and/or sbottoms (B, C, D) close in mass to the stops and neutralino respectively. The light sbottoms might allow the Bino DM to be a thermal relic by opening up a co-annihilation channel, and certainly remove overclosure bounds from the scenario, even for standard cosmological histories.

\begin{table}
\begin{center}
\begin{tabular}{| l || l | l || l | l | l |}
\hline
Scenario  & Explains \ww & Explains trilepton  & Natural SUSY & thermal  & $(g-2)_\mu$\\
& excess \cite{ATLAS:2012mec, CMS:2012tbb,CMS:2012daa}  & excess \cite{atlastrilepton} & spectrum & DM relic & \\
\hline 
\hline A & Yes & Yes & partial & No & No\\
\hline B & Yes & No & partial & possible & possible\\
\hline C & Yes & Yes & Yes & possible & No\\
\hline D & Yes & No & Yes & possible & possible\\ 
\hline
\end{tabular}
\end{center}
\caption{
Summarized phenomenological consequences of the four stop scenarios illustrated in \fref{scenarios}.  A thermal DM relic requires light sbottoms close to the Bino mass. Explaining $(g-2)_\mu$ requires sleptons to be inserted into the spectrum. See \sref{scenarios} for details.
}
\label{t.summary}
\end{table}

Scenarios A and C are particularly intriguing in light of the ATLAS trilepton search \cite{atlastrilepton}, which was expected to exclude much of the $WW$-preferred region of these scenarios but instead observes an excess which is precisely consistent with the spectrum required to explain the \ww excess. Since this signal is completely uncorrelated from the dilepton + MET final state of the \ww measurements, it lends additional weight to these scenarios as serious possibilities.

On the other hand, Scenarios B and D (without charginos) allow for the insertion of sleptons between the stop and LSP, which can help explain the $(g-2)_\mu$ anomaly, carrying over a desirable feature from the slepton \ww explanation in \cite{Curtin:2013gta}.

In the fully natural scenarios the Higgs coupling to gluons is expected to be $\sim 20-35\%$ larger than in the SM, with correspondingly enhanced signal strengths for gluon-initiated Higgs production modes. This is in some conflict with CMS measurements but somewhat favored by ATLAS, and also relies on the assumption that other coupling corrections are absent.   
Of course, it is also important to understand that any new EW scale physics could potentially contaminate Higgs search modes and change signal strengths from their SM values. There could also be additional shifts in the Higgs couplings from whatever particular mechanism generates the Higgs mass.

The fully natural SUSY explanation for the \ww excess therefore makes some universal predictions: stops near 220 GeV (which could be differentiated from the SM \ww signal by use of kinematic discriminants \cite{Rolbiecki:2013fia}), specific Higgs coupling corrections (if they act alone) and a possible trilepton chargino-neutralino signal which may already have been detected. The light sbottom near $\sim 130 \gev$ may also be detectable, if a fully inclusive search is performed where a highly squeezed sbottom decay does not fail some reconstruction requirement or veto. Even if we assume that the \ww excess is a fluctuation or some under-estimated systematic error in the SM prediction, it is a necessary consequence of that interpretation that the \ww signal region is poorly constrained, and as a result these fully natural SUSY spectra cannot be excluded at the present time. There is still hope  for naturalness.

\subsection*{Note}

Simultaneous to our work, ref.~\cite{Kim:2014eva} has also investigated the \ww excess within a subset of the SUSY models analyzed in this paper, and has come to similar results.

\subsection*{Acknowledgements}

We are very grateful to Matt Reece for helpful discussions about Higgs coupling constraints. 
The work of D.C. was supported in part by the National Science Foundation under Grant PHY-PHY-0969739.  The work of P.M.  and P.T.was supported in part by NSF CAREER Award NSF-PHY-1056833. 

\appendix

\section{Monte Carlo Simulation}
\label{a.montecarlo} \setcounter{equation}{0} \setcounter{footnote}{0}

This appendix outlines how we determined regions of the stop-neutralino mass plane that are preferred or excluded by the \ww cross section measurements \cite{ATLAS:2012mec, CMS:2012tbb, CMS:2012daa} in Scenario A and C (Figs.~\ref{f.stopww1} and~\ref{f.stopww2}).

For each mass plane, a grid of SLHA spectrum files with decay tables was created using \texttt{CPsuperH 2.3} \cite{cpsuperh}. Stop pair production was simulated at LO using using \texttt{Pythia 6.4/8.16}~\cite{pythia} (hard process/shower), and analyzed in a \texttt{FastJet 3.0.3}  \cite{fastjet} based analysis code. Our detector simulation takes into account lepton isolation requirements, experiment-specific identification efficiencies, and geometrical acceptances. Since we did not explicitly include detector effects we verified all distributions against the standard \texttt{MadGraph5(v2.1.1)} $\to$ \texttt{Pythia6} $\to$ \texttt{PGS} pipeline \cite{Alwall:2011uj, pythia}, and found no indication that our simulations were unreliable. (We corrected a bug in PGS to fix MET-smearing, but it did not affect our conclusions in this case.) All production was rescaled to NLO production cross sections calculated \texttt{Prospino 2.1}~\cite{prospino}. 

This procedure resulted in predictions for each Scenario's contribution to the various kinematic distributions shown in the \ww cross section measurements. This allowed us to define a $\chi^2$ function for each point in each Scenario's stop-neutralino mass plane in the following fashion:
\begin{equation}
\chi^2(r_{SM}, r_{BSM}; m_{\tilde t}, m_{\tilde \chi^0_1})
\end{equation}
which was obtained by comparing to experimental data the predicted SM contributions in all kinematic distributions from \cite{ATLAS:2012mec, CMS:2012tbb, CMS:2012daa}, normalized by factor $r_{SM}$, with the added BSM contribution at the respective scenario's mass point $(m_{\tilde t}, m_{\tilde \chi^0_1})$, normalized by factor $r_{BSM}$.

We then defined a $\chi^2$ ratio
\begin{equation}
\frac{\chi^2(1, 1; m_{\tilde t}, m_{\tilde \chi^0_1})}{\chi^2(1, 0)}
\end{equation}
to evaluate how much the stop contribution improved $(< 1)$ or degraded $(>1)$ agreement with data compared to the SM at each mass point. This gave the light blue contours and the \ww preferred regions in Figs.~\ref{f.stopww1} and~\ref{f.stopww2}. Since the experiments do not make full likelihoods available the specific values we obtain for the $\chi^2$ are not exactly correct, but the qualitative statement that certain regions of the mass plane are preferred should be robust.

Exclusions on the stop scenarios were obtained from the \ww measurements in two ways. To be conservative, one could decide not to trust the SM prediction for the total \ww cross section. In this case, we defined the best-fit $\chi^2$ for each point by minimizing with respect to $r_{SM}$:
\begin{equation}
\chi^2_\mathrm{float}(m_{\tilde t}, m_{\tilde \chi^0_1})  \ \ \equiv \ \  \underset{r_{SM}}{\mathrm{min}} \  \chi^2(r_{SM}, 1; m_{\tilde t},m_{\tilde \chi^0_1}).
\end{equation}
Stronger exclusions can be obtained by trusting the normalization of the SM contribtions. In that case we simply define
\begin{equation}
\chi^2_\mathrm{fixed}(m_{\tilde t}, m_{\tilde \chi^0_1})  \ \ \equiv \ \   \chi^2(1, 1; m_{\tilde t},m_{\tilde \chi^0_1}).
\end{equation}
Contours where $\chi^2_\mathrm{float}(m_{\tilde t}, m_{\tilde \chi^0_1})$  and $\chi^2_\mathrm{fixed}(m_{\tilde t}, m_{\tilde \chi^0_1})$  gave a $p$-value of 0.05 are given as $95\%$ CL exclusions (solid and dashed orange lines) in Figs.~\ref{f.stopww1} and~\ref{f.stopww2}. The bound obtained with floating SM contribution should be very robust even in light of possible future corrections to the SM \ww cross section calculation, unless they significantly change the expected shape of kinematic distributions.

The sbottom bounds in all figures could be applied to our scenarios directly. The same was true of the stop bounds (with the exception of some simple rescaling for \fref{stopww2}), except for the light stop search \cite{ATLAS7stop1} which looked for the correct final state but did not supply exclusions for the specific squeezed spectra $m_{\tilde t} - m_{\tilde \chi^\pm_1} \lesssim 10 \gev$ featured in our scenarios. We recast this search by implementing the corresponding cuts in our simulation scheme, rescaling our acceptances by 0.5 to match the expected BSM acceptances they supply, and obtaining exclusions from the number of events they observe in each signal bin.



\begin{thebibliography}{99}

 \bibitem{Higgs125Combined} 
 
 ATLAS-CONF-2013-034 (http://cds.cern.ch/record/1528170);
 
  G.~Aad {\it et al.}  [ATLAS Collaboration],
  Phys.\ Lett.\ B {\bf 716}, 1 (2012)
  [arXiv:1207.7214 [hep-ex]];
  
  S.~Chatrchyan {\it et al.}  [CMS Collaboration],
  Phys.\ Lett.\ B {\bf 716}, 30 (2012)
  [arXiv:1207.7235 [hep-ex]].
 


  \bibitem{moriondEW}
  Talks given at March 2014 Electroweak Moriond Conference, specifically:
  ``Strong SUSY production searches'' (Pedrame Bargassa), ``EW SUSY production searches at ATLAS and CMS'' (Michael Flowerdew), ``Multilepton and Multiphoton signatures of SUSY at the LHC'' (Christoffer Petersson), ``Dark Matter searches in LHC'' (Philippe Calfayan), ``Exotic Searches in LHC'' (Thiago Rafael).
  
  \bibitem{moriondQCD}
  Talks given at March 2014 QCD Moriond Conference, specifically:
  ``Searches for BSM Higgs Bosons at the LHC'' (Paolo Meridiani), ``Third Generation SUSY Searches at the LHC'' (Takashi Yamanaka), ``Inclusive SUSY Searches at the LHC'' (Sezen Sekmen), ``Searches for Heavy Resonances at the LHC'' (Tetiana Hryn'Ova), ``Searches for dark matter and extra dimensions at the LHC'' (Sarah Eno).
  
  
  
  
  



\bibitem{nmssm}
  J.~R.~Espinosa and M.~Quiros,
  Phys.\ Lett.\ B {\bf 279}, 92 (1992).
  U.~Ellwanger, C.~Hugonie and A.~M.~Teixeira,
  Phys.\ Rept.\  {\bf 496}, 1 (2010)
  [arXiv:0910.1785 [hep-ph]].

\bibitem{lambdasusy} 
  L.~J.~Hall, D.~Pinner and J.~T.~Ruderman,
  JHEP {\bf 1204}, 131 (2012)
  [arXiv:1112.2703 [hep-ph]].

\bibitem{dterms} 
  P.~Batra, A.~Delgado, D.~E.~Kaplan and T.~M.~P.~Tait,
  JHEP {\bf 0402}, 043 (2004)
  [hep-ph/0309149].
  
  
\bibitem{stoptotopsearches}
CMS SUS-13-004, CMS SUS-13-011, ATLAS-CONF-2013-024, ATLAS-CONF-2013-037.


\bibitem{otherstopsearchesATLAS}
  The ATLAS collaboration,
  ATLAS-CONF-2013-054;
   The ATLAS collaboration,
  ATLAS-CONF-2013-061;
[ATLAS Collaboration],
  ATLAS-CONF-2012-151;
   [ATLAS Collaboration],
  ATLAS-CONF-2013-007;
  [ATLAS Collaboration],
  ATLAS-CONF-2013-037;
  The ATLAS collaboration,
  ATLAS-CONF-2013-053;
  G.~Aad {\it et al.}  [ATLAS Collaboration],
  Phys.\ Lett.\ B {\bf 720}, 13 (2013)
  [arXiv:1209.2102 [hep-ex]];
  G.~Aad {\it et al.}  [ATLAS Collaboration],
  Phys.\ Rev.\ Lett.\  {\bf 109}, 211802 (2012)
  [arXiv:1208.1447 [hep-ex]];
  G.~Aad {\it et al.}  [ATLAS Collaboration],
  Phys.\ Rev.\ Lett.\  {\bf 109}, 211803 (2012)
  [arXiv:1208.2590 [hep-ex]];
  G.~Aad {\it et al.}  [ATLAS Collaboration],
  JHEP {\bf 1211}, 094 (2012)
  [arXiv:1209.4186 [hep-ex]].
  
  
  
  
  
  
  
  

\bibitem{otherstopsearchesCMS}
  S.~Chatrchyan {\it et al.}  [CMS Collaboration],
  arXiv:1402.4770 [hep-ex];
  S.~Chatrchyan {\it et al.}  [CMS Collaboration],
  arXiv:1311.4937 [hep-ex];
  CMS Collaboration [CMS Collaboration],
  CMS-PAS-SUS-13-016;
  CMS Collaboration [CMS Collaboration],
  CMS-PAS-SUS-13-013;
  CMS Collaboration [CMS Collaboration],
  CMS-PAS-SUS-13-008.










\bibitem{stoptocharmsearches} 
  CMS Collaboration [CMS Collaboration],
  CMS-PAS-SUS-13-009;
    The ATLAS collaboration,
  ATLAS-CONF-2013-068.
  
  
  \bibitem{Aad:2014qaa} 
  G.~Aad {\it et al.}  [ATLAS Collaboration],
  arXiv:1403.4853 [hep-ex].
  

  
 \bibitem{ATLAS7stop1}
  [ATLAS Collaboration],[arXiv:1208.4305 [hep-ex]]

\bibitem{generictoppartnersearches}
  [ATLAS Collaboration],
  ATLAS-CONF-2013-018;
  G.~Aad {\it et al.}  [ATLAS Collaboration],
  Phys.\ Lett.\ B {\bf 718}, 1284 (2013)
  [arXiv:1210.5468 [hep-ex]];
  S.~Chatrchyan {\it et al.}  [CMS Collaboration],
  Phys.\ Lett.\ B {\bf 716}, 103 (2012)
  [arXiv:1203.5410 [hep-ex]].


\bibitem{meadereece} 
  P.~Meade and M.~Reece,
  Phys.\ Rev.\ D {\bf 74}, 015010 (2006)
  [hep-ph/0601124].

\bibitem{tophunter} 
  A.~De Simone, O.~Matsedonskyi, R.~Rattazzi and A.~Wulzer,
  JHEP {\bf 1304}, 004 (2013)
  [arXiv:1211.5663 [hep-ph]].

 
 


\bibitem{hidinglightstops}
  Y.~Kats and D.~Shih,
  JHEP {\bf 1108}, 049 (2011)
  [arXiv:1106.0030 [hep-ph]];
   C.~Brust, A.~Katz and R.~Sundrum,
  JHEP {\bf 1208}, 059 (2012)
  [arXiv:1206.2353 [hep-ph]];
   Z.~Han, A.~Katz, D.~Krohn and M.~Reece,
  JHEP {\bf 1208}, 083 (2012)
  [arXiv:1205.5808 [hep-ph]];
    C.~Brust, A.~Katz, S.~Lawrence and R.~Sundrum,
  JHEP {\bf 1203}, 103 (2012)
  [arXiv:1110.6670 [hep-ph]];
    J.~A.~Evans and Y.~Kats,
  JHEP {\bf 1304}, 028 (2013)
  [arXiv:1209.0764 [hep-ph]].


\bibitem{Curtin:2012nn} 
  D.~Curtin, P.~Jaiswal and P.~Meade,
 Phys.\ Rev.\ D {\bf 87}, 031701(R) (2013)
 [arXiv:1206.6888 [hep-ph]].

\bibitem{ATLAS:2012mec} 
  G.~Aad {\it et al.}  [ATLAS Collaboration],
  Phys.\ Rev.\ D {\bf 87}, no. 11, 112001 (2013)
  [Erratum-ibid.\ D {\bf 88}, no. 7, 079906 (2013)]
  [arXiv:1210.2979 [hep-ex]].
  

\bibitem{CMS:2012tbb} 
  CMS Collaboration [CMS Collaboration],
  CMS-PAS-SMP-12-005.
  
  
\bibitem{CMS:2012daa} 
  CMS Collaboration [CMS Collaboration],
  CMS-PAS-SMP-12-013.
  
  

\bibitem{atlashwwdetailed}
[ATLAS Collaboration],
  ATLAS-CONF-2013-030.

\bibitem{Rolbiecki:2013fia} 
  K.~Rolbiecki and K.~Sakurai,
  JHEP {\bf 1309}, 004 (2013)
  [arXiv:1303.5696 [hep-ph]].
  

 \bibitem{Curtin:2013gta} 
  D.~Curtin, P.~Jaiswal, P.~Meade and P.~-J.~Tien,
  JHEP {\bf 1308}, 068 (2013)
  [arXiv:1304.7011 [hep-ph]].
 
 
 

\bibitem{comphiggsreview} 
  R.~Contino,
  arXiv:1005.4269 [hep-ph];
  B.~Bellazzini, C.~Csaki and J.~Serra,
  arXiv:1401.2457 [hep-ph].


  
\bibitem{ewbghiggs} 
  D.~Curtin, P.~Jaiswal and P.~Meade,
  JHEP {\bf 1208}, 005 (2012)
  [arXiv:1203.2932 [hep-ph]];
   T.~Cohen, D.~E.~Morrissey and A.~Pierce,
  Phys.\ Rev.\ D {\bf 86}, 013009 (2012)
  [arXiv:1203.2924 [hep-ph]];
    T.~Cohen and A.~Pierce,
  Phys.\ Rev.\ D {\bf 85}, 033006 (2012)
  [arXiv:1110.0482 [hep-ph]].


\bibitem{Fan:2014txa} 
  J.~Fan and M.~Reece,
  arXiv:1401.7671 [hep-ph].
  
  
\bibitem{Carena:2013iba} 
  M.~Carena, S.~Gori, N.~R.~Shah, C.~E.~M.~Wagner and L.~-T.~Wang,
  JHEP {\bf 1308}, 087 (2013)
  [arXiv:1303.4414, arXiv:1303.4414 [hep-ph]].











\bibitem{mattemail}
Private email communication with Matt Reece, March 6 2014.

  
  
  \bibitem{atlastrilepton} 
 G.~Aad {\it et al.}  [ATLAS Collaboration],
  JHEP {\bf 1404}, 169 (2014)
  [arXiv:1402.7029 [hep-ex]];
   G.~Aad {\it et al.}  [ATLAS Collaboration],
  arXiv:1403.5294 [hep-ex].
  

\bibitem{cmstrilepton} 
  V.~Khachatryan {\it et al.}  [ CMS Collaboration],
  arXiv:1405.7570 [hep-ex].
  
  
  \bibitem{CMS:2013afa} 
  CMS Collaboration [CMS Collaboration],
  CMS-PAS-SUS-13-017.
  
  
  \bibitem{TheATLAScollaboration:2013zia} 
  The ATLAS collaboration,
  ATLAS-CONF-2013-093.
  
  
  
  \bibitem{Kribs:2008hq} 
  G.~D.~Kribs, A.~Martin and T.~S.~Roy,
  JHEP {\bf 0901}, 023 (2009)
  [arXiv:0807.4936 [hep-ph]].
  
  
 
\bibitem{ATLAScharginotoWMETdileptonsearch} 
  The ATLAS collaboration,
  ATLAS-CONF-2013-049.
  


 \bibitem{ATLASsb2}
 ATLAS-CONF-2012-165


\bibitem{obliquesusy}
  S.~P.~Martin, K.~Tobe and J.~D.~Wells,
  Phys.\ Rev.\ D {\bf 71} (2005) 073014
  [hep-ph/0412424].
  
  \bibitem{bsgamma} 
  A.~Katz, M.~Reece and A.~Sajjad,
  arXiv:1406.1172 [hep-ph].




 \bibitem{ATLAS8stop1}
ATLAS-CONF-2012-167


 \bibitem{LEPsb}
LEPSUSYWG: ALEPH, DELPHI, L3, and OPAL Collaborations,
 , http://lepsusy.web.cern.ch/lepsusy, report No. LEPSUSYWG/04-02.1..


\bibitem{D0sb}
  [D0 Collaboration],
Phys.\ Lett.\ B {\bf 693}, 95 (2010);
[arXiv:1005.2222 [hep-ex]]

 \bibitem{Yangbai}
Ezequiel Alvarez and Yang Bai,
  [arXiv:1204.5182 [hep-ph]].

  \bibitem{Blum:2012ii} 
  K.~Blum, R.~T.~D'Agnolo and J.~Fan,
  JHEP {\bf 1301}, 057 (2013)
  [arXiv:1206.5303 [hep-ph]].  
  
  
  \bibitem{welltempered}
  N.~Arkani-Hamed, A.~Delgado and G.~F.~Giudice,
  Nucl.\ Phys.\ B {\bf 741}, 108 (2006)
  [hep-ph/0601041].
  
 \bibitem{planck} 
  P.~A.~R.~Ade {\it et al.}  [Planck Collaboration],
  arXiv:1303.5076 [astro-ph.CO].
 
 
 \bibitem{micromegas}
  G.~Belanger, F.~Boudjema, A.~Pukhov and A.~Semenov,
  Comput.\ Phys.\ Commun.\  {\bf 176}, 367 (2007)
  [hep-ph/0607059].



 \bibitem{luxdm}
First results from the LUX dark matter experiment at the Sanford Underground Research Facility;
[arXiv:1310.8214v2[astro-ph.CO]]   
    
    

   \bibitem{pdg:2012} 
  pdg,
 http://pdg.lbl.gov/2012/reviews/rpp2012-rev-g-2-muon-anom-mag-moment.pdf
 
 \bibitem{djouadi2} 
A.~Djouadi,
Phys.\ Rept.\  {\bf 459}, 1 (2008)
[hep-ph/0503173].


\bibitem{CMS:ril} 
  [CMS Collaboration],
  CMS-PAS-HIG-13-001.
  
  
  
  

\bibitem{ATLAS:2013oma} 
  [ATLAS Collaboration],
  ATLAS-CONF-2013-012.
  
  
\bibitem{Dawson:2013lya} 
  S.~Dawson, I.~M.~Lewis and M.~Zeng,
  Phys.\ Rev.\ D {\bf 88}, no. 5, 054028 (2013)
  [arXiv:1307.3249].
 
  
\bibitem{ZZmeasurement} 
  [ATLAS Collaboration],
  ATLAS-CONF-2013-020;
   CMS Collaboration [CMS Collaboration],
gauge couplings in lll'l' decays at sqrt(s) = 8 TeV at the LHC,''
  CMS-PAS-SMP-13-005.
  
  
 \bibitem{zznnlo} 
  F.~Cascioli, T.~Gehrmann, M.~Grazzini, S.~Kallweit, P.~Maierhšfer, A.~von Manteuffel, S.~Pozzorini and D.~Rathlev {\it et al.},
  arXiv:1405.2219 [hep-ph].
    
    
  
  \bibitem{resummation} 
  P.~Meade, H.~Ramani, M.~Zeng, to appear shortly.


  
 \bibitem{Kim:2014eva} 
  J.~S.~Kim, K.~Rolbiecki, K.~Sakurai and J.~Tattersall,
  arXiv:1406.0858 [hep-ph].

\bibitem{cpsuperh}
J.~S.~Lee, A.~Pilaftsis, M.~S.~Carena, S.~Y.~Choi, M.~Drees, J.~R.~Ellis and C.~E.~M.~Wagner,
  Comput.\ Phys.\ Commun.\  {\bf 156}, 283 (2004)
  [hep-ph/0307377];
  J.~S.~Lee, M.~Carena, J.~Ellis, A.~Pilaftsis and C.~E.~M.~Wagner,
  Comput.\ Phys.\ Commun.\  {\bf 180}, 312 (2009)
  [arXiv:0712.2360 [hep-ph]];
    J.~S.~Lee, M.~Carena, J.~Ellis, A.~Pilaftsis and C.~E.~M.~Wagner,
  Comput.\ Phys.\ Commun.\  {\bf 184}, 1220 (2013)
  [arXiv:1208.2212 [hep-ph]].
  
  
  
\bibitem{pythia} 
  T.~Sjostrand, S.~Mrenna and P.~Z.~Skands,
  Comput.\ Phys.\ Commun.\  {\bf 178}, 852 (2008)
  [arXiv:0710.3820 [hep-ph]].
    T.~Sjostrand, S.~Mrenna and P.~Z.~Skands,
  JHEP {\bf 0605}, 026 (2006)
  [hep-ph/0603175].
  
    \bibitem{fastjet}
    M.~Cacciari and G.~P.~Salam,
  Phys.\ Lett.\ B {\bf 641}, 57 (2006)
  [hep-ph/0512210];
    M.~Cacciari, G.~P.~Salam and G.~Soyez,
  Eur.\ Phys.\ J.\ C {\bf 72}, 1896 (2012)
  [arXiv:1111.6097 [hep-ph]].


\bibitem{Alwall:2011uj} 
  J.~Alwall, M.~Herquet, F.~Maltoni, O.~Mattelaer and T.~Stelzer,
  JHEP {\bf 1106}, 128 (2011)
  [arXiv:1106.0522 [hep-ph]].
  
  
  \bibitem{prospino}
  W.~Beenakker, R.~Hopker and M.~Spira,
  hep-ph/9611232.
    W.~Beenakker, M.~Klasen, M.~Kramer, T.~Plehn, M.~Spira and P.~M.~Zerwas,
  Phys.\ Rev.\ Lett.\  {\bf 83}, 3780 (1999)
  [Erratum-ibid.\  {\bf 100}, 029901 (2008)]
  [hep-ph/9906298].

 \end{thebibliography}
\end{document}